\documentstyle[aps,prb,multicol]{revtex}
\begin{document}
\title{
Exact Drude weight for the one-dimensional Hubbard model\\
at finite temperatures
}
\author{Satoshi Fujimoto}
\address{
Department of Physics, Theoretical Physics,
University of Oxford, 1Keble Road, Oxford,
OX1 3NP, U.K. \\
Department of Physics,
Kyoto University, Kyoto 606, Japan
}
\author{Norio Kawakami}
\address{
Department of Applied Physics, 
Osaka University, Suita, Osaka 565, Japan
}
\date{\today}

\maketitle

\begin{abstract}
The Drude weight for the one-dimensional
Hubbard model is investigated at finite temperatures
by using the Bethe ansatz solution.
Evaluating finite-size corrections to the thermodynamic Bethe ansatz
equations, we obtain the formula for  the Drude weight 
as the response of the system to an external gauge potential.
We perform low-temperature expansions of the
Drude weight in the case of half-filling as well as
away from half-filling, which 
clearly distinguish the Mott-insulating state from 
the metallic state.
\end{abstract}

\pacs{PACS numbers: }


\section{Introduction}

The Mott-Hubbard metal-insulator transition (MIT) is one of
the long-standing important issues in strongly correlated electron
systems. The one-dimensional (1D) Hubbard model is a fundamental model
which describes the MIT.  This model is exactly solvable in terms of 
the Bethe ansatz method\cite{lw}.
Various thermodynamic quantities 
such as the specific heat and the spin- and charge-susceptibilities 
which characterize the MIT have been 
obtained exactly\cite{taka1,taka2,kawa1}.
However, the study on transport properties such as the Drude weight 
based upon the Bethe ansatz method 
was restricted to the zero temperature case\cite{shastry,schulz,kaya}.
The Drude weight at finite temperatures
was investigated only by using numerical methods 
so far\cite{zotos,zotos2}.
Since the Drude weight is a direct probe for the MIT\cite{kohn,shastry}, 
it is desirable to obtain the exact results on it 
at finite temperatures. This is the main purpose of this paper. 
In order to obtain the Drude weight, it is necessary to calculate the
finite-size corrections to the energy spectrum,
so that we should evaluate the finite-size corrections to
the thermodynamic Bethe ansatz equations.
For this purpose, we generalize standard methods 
for finite-size corrections based on the Euler-Maclaurin formula
to the case of finite temperatures, and obtain the leading temperature 
dependence of the Drude weight at low temperatures.

It is easily seen from the Kubo formula that 
the Drude weight in translationally invariant systems 
is independent of temperature.
Then the temperature dependence of the Drude weight for the Hubbard
model comes from  the interactions such as the Umklapp scattering
which break translational symmetry.  This implies that
its temperature dependence carries the information
how the electron correlations which control the MIT develop
in low temperatures.
In the case of half-filling, 
we indeed show how the system approaches the Mott insulator
as the temperature is decreased.

The organization of this paper is as follows.
In Sec. II, the formulation 
for the finite-size corrections to thermodynamic Bethe ansatz
equations is given.  The procedure is not restricted to the 
Hubbard model, but also applicable to any other solvable models.
In Sec. III, we obtain the general expression for the Drude 
weight at finite temperatures.
The low-temperature expansion for the Drude weight is derived
in Sec. IV  both in the case of  half-filling as well as 
away from half-filling. The leading temperature dependence is 
evaluated  at low temperatures. 
 Brief summary is  given in Sec. V.

\section{Finite-size corrections to the thermodynamic Bethe ansatz
equations}

In this section we consider finite-size effects on the Bethe ansatz
solutions of the 1D Hubbard model at finite temperatures.
Finite-size effects at zero temperature have been
studied by many authors in connection with the application of 
conformal field theory\cite{we,woy,ky,fk}.
We generalize their method to the case of finite temperatures.
The hamiltonian of the 1D Hubbard model reads,
\begin{equation}
H=-\sum_{\sigma i}c^{\dagger}_{\sigma i}c_{\sigma i+1}+h.c.
+U\sum_i c^{\dagger}_{\uparrow i}c_{\uparrow i}
c^{\dagger}_{\downarrow i}c_{\downarrow i}.
\end{equation}
It is necessary to introduce the 
Aharonov-Bohm (AB) flux $\Phi$ to formulate
the Drude weight as the response to an external gauge
potential. Alternatively, the effect of the AB flux is 
incorporated into the twisted boundary condition for the wave function,
$\Psi(x+L)=e^{i\Phi}\Psi(x)$\cite{by}.
The Bethe ansatz equations in the presence of the AB flux
are given by\cite{lw,ss},
\begin{equation}
e^{ik_jL}=e^{i\Phi}\prod_{\alpha=1}^{M}
\frac{\sin k_j-\Lambda_{\alpha}+iu}{\sin k_j-\Lambda_{\alpha}-iu},
\label{be1}
\end{equation}
\begin{equation}
\prod_{j=1}^{N}\frac{\Lambda_{\alpha}-\sin k_j+iu}
{\Lambda_{\alpha}-\sin k_j-iu}
=-\prod_{\beta=1}^{M}\frac{\Lambda_{\alpha}-\Lambda_{\beta}+2iu}
{\Lambda_{\alpha}-\Lambda_{\beta}-2iu}.\label{be2}
\end{equation}
Here $k_j$ and $\Lambda_{\alpha}$ are, respectively, the rapidities 
for the charge and spin degrees of freedom, and we have introduced
 $u=U/4$. $N$ and $M$ are the total 
number of electrons and down spins.

Thermodynamic Bethe ansatz solution for the 1D Hubbard model
was obtained by Takahashi many years ago with the use of
so-called string hypothesis\cite{taka1}.
The validity of the string hypothesis is justified only for 
the thermodynamic limit $L\rightarrow \infty$.
In order to calculate the Drude weight, we should
evaluate the energy for a finite-size system,
because the effect of $\Phi$ vanishes in the thermodynamic limit.
Thus one may worry about whether the string hypothesis can be applied 
to the calculation of the Drude weight. 
However, by recalling the following fact 
we can still adopt the string hypothesis for our purpose:
The corrections to the string hypothesis for a finite-size 
system is estimated as $\sim O(e^{-cL})$, where $c$ is a  constant
which depends on the temperature.
On the other hand, the dependence  of the energy on the AB flux
appears in the order of  $ 1/L^2$.
Thus the correction to the string hypothesis 
for the finite-size system
is much smaller than the finite-size corrections
to the energy spectrum which we need for the calculation
of the Drude weight. This situation is analogous to that for  
the Kondo model\cite{andrei}  or the impurity Anderson mdel
\cite{wieg,kawa2},  where the local electron correlations, which are 
given by the $1/L$-corrections
to bulk quantities, are correctly evaluated based upon 
the string hypothesis. 

Using the string hypothesis, we can thus 
write down  the thermodynamic
Bethe ansatz equations. After  taking the logarithm of 
the above equations, we end up with
\begin{eqnarray}
k_jL&=&2\pi I_j+\Phi-\sum_{n=1}^{\infty}\sum_{\alpha=1}^{M_n}
\theta\biggl(\frac{\sin k_j-\Lambda_{\alpha}^{n}}{nu}\biggr) 
\nonumber \\
&&-\sum_{n=1}^{\infty}\sum_{\alpha=1}^{M_n'}
\theta\biggl(\frac{\sin k_j-{\Lambda'}_{\alpha}^{n}}{nu}
\biggr), 
\label{bet1}
\end{eqnarray}
\begin{eqnarray}
&&L(\sin^{-1}({\Lambda'}_{\alpha}^{n}+inu)
+\sin^{-1}({\Lambda'}_{\alpha}^{n}-inu))
=2\pi {J'}_{\alpha}^{n}+2n\Phi \nonumber \\
&+&\sum_{j=1}^{N-2M'}
\theta\biggl(\frac{\sin k_j-{\Lambda'}_{\alpha}^{n}}{nu}
\biggr) 
+\sum_{m=1}^{\infty}\sum_{\beta}
\Theta_{nm}\biggl(
\frac{{\Lambda'}_{\alpha}^{n}-{\Lambda'}_{\beta}^{m}}{u}\biggr),
\label{bet2}
\end{eqnarray}
\begin{equation}
\sum_{j=1}^{N-2M'}
\theta\biggl(
\frac{\sin k_j-\Lambda_{\alpha}^{n}}{nu}\biggr)
=2\pi J_{\alpha}^n +\sum_{m=1}^{\infty}\sum_{\beta}
\Theta_{nm}\biggl(
\frac{\Lambda_{\alpha}^{n}-\Lambda_{\beta}^{m}}{u}\biggr),
\label{bet3}
\end{equation}
with $\theta(x)=2\tan^{-1}x$ and 
\begin{equation}
\Theta_{nm}(x)=\left\{
\begin{array}{ll}
\displaystyle{\theta\biggl(\frac{x}{\vert n-m\vert}\biggr)
+2\theta\biggl(\frac{x}{\vert n-m\vert +2}\biggr)
+\cdot\cdot\cdot
+2\theta\biggl(\frac{x}{ n+m-2}\biggr)
+\theta\biggl(\frac{x}{n+m}\biggr)}&\qquad n\neq m \\
\displaystyle{2\theta\biggl(\frac{x}{2}\biggr)+\cdot\cdot\cdot
+2\theta\biggl(
\frac{x}{2n-2}\biggr)+\theta\biggl(\frac{x}{2n}\biggr)}
&\qquad n=m.
\end{array}
\right. 
\end{equation}
Here $k_j$ is the rapidity for charge excitations which
are not in bound states, 
$\Lambda_{\alpha}^{n}$ is that for spin excitations,
and ${\Lambda'}_{\alpha}^{n}$ is that for 
bound states.
$I_j$, $J_{\alpha}^{n}$, and ${J'}_{\alpha}^{n}$ are the corresponding 
quantum numbers which specify the above excitations, respectively.
$M_n$ is the number of $n$-strings for spin excitations.
$2M'$ is the total number of electrons which make
bound states.

In order to calculate the finite-size corrections to
the energy spectrum, we expand the rapidities in terms of $1/L$
following Berkovich and Murthy\cite{bm},
\begin{eqnarray}
k_j&=&k_j^{\infty}+\frac{f_{1}}{L}+\frac{f_2}{L^2}
+O(1/L^3), \nonumber \\ 
\Lambda_{\alpha}^{n}&=&\Lambda_{\alpha}^{n\infty}
+\frac{g_{1n}}{L}+\frac{g_{2n}}{L^2}+O(1/L^3), \nonumber \\
{\Lambda'}_{\alpha}^{n}&=&{\Lambda'}_{\alpha}^{n\infty}
+\frac{h_{1n}}{L}+\frac{h_{2n}}{L^2}+O(1/L^3). \label{rap}
\end{eqnarray}
The lowest-order contributions in $1/L$ 
give the conventional thermodynamic Bethe ansatz equations
which read\cite{taka1},
\begin{equation}
(1+\zeta(k))\rho(k)=\frac{1}{2\pi}+\sum_{n=1}^{\infty}
\int^{\infty}_{-\infty}\frac{d\Lambda}{\pi}
\frac{nu\cos k(\sigma_n(\Lambda)+{\sigma'}_n(\Lambda))}
{(nu)^2+(\sin k-\Lambda)^2},
\label{tba1}
\end{equation}
\begin{equation}
\eta_n(\Lambda)\sigma_n(\Lambda)+\sum_{m=1}^{\infty}
A_{nm}*\sigma_m(\Lambda)=
\int^{\infty}_{\infty}\frac{dk}{\pi}\frac{nu\rho(k)}
{(nu)^2+(\sin k-\Lambda)^2},
\label{tba2}
\end{equation}
\begin{equation}
{\eta'}_n(\Lambda){\sigma'}_n(\Lambda)+\sum_{m=1}^{\infty}
A_{nm}*{\sigma'}_m(\Lambda)=
\frac{1}{\pi}{\rm Re}\frac{1}{\sqrt{1-(\Lambda-inu)^2}}
-\int^{\infty}_{\infty}\frac{dk}{\pi}\frac{nu\rho(k)}
{(nu)^2+(\sin k-\Lambda)^2},
\label{tba3}
\end{equation}
\begin{eqnarray}
\ln\zeta(k)=\frac{-2\cos k -\mu_0 H-A}{T} 
+\sum_{n=1}^{\infty}\int^{\infty}_{-\infty}\frac{d\Lambda}{\pi}
\frac{nu(\ln(1+{\eta'}_n^{-1}(\Lambda))
-\ln(1+\eta_n^{-1}(\Lambda)))}{(nu)^2+(\sin k-\Lambda)^2},
\label{tba4}
\end{eqnarray}
\begin{eqnarray}
\ln(1+\eta_n(\Lambda))+\int^{\pi}_{-\pi}
\frac{dk}{\pi}\frac{nu\cos k \ln(1+\zeta^{-1}(k))}
{(nu)^2+(\sin k-\Lambda)^2} 
=\frac{2n\mu_0H}{T}+\sum_{m=1}^{\infty}A_{nm}*
\ln(1+\eta_m^{-1}(\Lambda)),
\label{tba5}
\end{eqnarray}
\begin{eqnarray}
\ln(1+{\eta'}_n(\Lambda))&+&\int^{\pi}_{-\pi}
\frac{dk}{\pi}\frac{nu\cos k \ln(1+\zeta^{-1}(k))}
{(nu)^2+(\sin k-\Lambda)^2} \nonumber \\ 
&=&\frac{4{\rm Re}\sqrt{1-(\Lambda-inu)^2}-2nA}{T}
+\sum_{m=1}^{\infty}A_{nm}*
\ln(1+{\eta'}_m^{-1}(\Lambda)),
\label{tba6}
\end{eqnarray}
where 
$$A_{nm}*\phi(x)=\delta_{nm}\phi(x)+\frac{d}{dx}
\int^{\infty}_{-\infty}\frac{dx'}{2\pi}
\Theta_{nm}(\frac{x-x'}{u})\phi(x').$$
$\rho(k)$, $\sigma_n(\Lambda)$, and ${\sigma'}_n(\Lambda)$
are the distribution functions for the rapidities, $k_j$, 
$\Lambda_{\alpha}^{n}$, and ${\Lambda'}_{\alpha}^{n}$,
respectively, and $\zeta(k)\equiv\rho^h/\rho$, 
$\eta_n(\Lambda)\equiv\sigma^h_n/\sigma_n$, 
${\eta'}_n(\Lambda)\equiv{\sigma'}_n^h/{\sigma'}_n$
with the distribution functions for holes
$\rho^h$, $\sigma^h_n$, and ${\sigma'}_n^h$. 
We have introduced an external magnetic field $H$ and a chemical
potential $A$.

Using the Euler-Maclaurin formula and eqs.(\ref{bet1})-(\ref{bet3})
and (\ref{rap}), we obtain the $1/L$- and $1/L^2$-corrections
to the Bethe ansatz equations which determine
$f_{1,2}$, $g_{1,2n}$ and $h_{1,2n}$.
Taking a continuum limit, we consequently have,
\begin{equation}
(1+\zeta(k))\rho(k)f_1(k)=\frac{\Phi}{2\pi}
+\sum_{n=1}^{\infty}\int^{\infty}_{-\infty} \frac{d\Lambda}{\pi}
\frac{nu}{(nu)^2+(\sin k-\Lambda)^2}(g_{1n}(\Lambda)
\sigma_{n}(\Lambda)+h_{1n}(\Lambda){\sigma'}_{n}(\Lambda)),
\label{f1}
\end{equation}
\begin{equation}
\sigma_n(\Lambda)\eta_n(\Lambda)g_{1n}(\Lambda)+
\sum_{m=1}^{\infty}A_{nm}* \sigma_m(\Lambda)g_{1n}(\Lambda)
=\int^{\pi}_{-\pi}\frac{dk}{\pi}\frac{nu\cos k}{(nu)^2+(\sin k-\Lambda)^2}
f_1(k)\rho(k),
\label{g1}
\end{equation}
\begin{equation}
{\sigma'}_n(\Lambda){\eta'}_n(\Lambda)h_{1n}(\Lambda)+
\sum_{m=1}^{\infty}A_{nm}* {\sigma'}_m(\Lambda)h_{1n}(\Lambda)
=\frac{n\Phi}{\pi}
-\int^{\pi}_{-\pi}\frac{dk}{\pi}\frac{nu\cos k}{(nu)^2+(\sin k-\Lambda)^2}
f_1(k)\rho(k),
\label{h1}
\end{equation}
\begin{eqnarray}
(1&+&\zeta(k))\rho(k)f_2(k)=\sum_{n=1}^{\infty}\int^{\infty}_{-\infty}
\frac{d\Lambda}{\pi}\frac{nu}{(nu)^2+(\sin k-\Lambda)^2}
(g_{2n}(\Lambda)\sigma_{n}(\Lambda)+h_{2n}(\Lambda){\sigma'}_n
(\Lambda) \nonumber \\
&&+\frac{1}{2}\frac{d}{dk}[(1+\zeta(k))\rho(k)f_1^2(k)] \nonumber \\
&&+\sum_{m=1}^{\infty}\int^{\infty}_{-\infty}\frac{d\Lambda}{\pi}
\frac{nu(\sin k-\Lambda)}{((nu)^2+(\sin k-\Lambda)^2)^2}
(g_{1n}^2(\Lambda)\sigma_n(\Lambda)+
h_{1n}^2(\Lambda){\sigma'}_n(\Lambda)),
\label{f2}
\end{eqnarray}
\begin{eqnarray}
(1&+&\eta_n(\Lambda))\sigma_n(\Lambda)g_{2n}(\Lambda)=
\frac{1}{2}\frac{d}{d\Lambda}[(1+\eta_n(\Lambda))\sigma_n(\Lambda)
g_{1n}^2(\Lambda)] \nonumber \\
&+&\int^{\pi}_{-\pi}\frac{dk}{\pi}\frac{nu\cos k}
{(nu)^2+(\sin k-\Lambda)^2}f_2(k)\rho(k)
+\int^{\pi}_{-\pi}\frac{dk}{\pi}\frac{nu(\sin k-\Lambda)\cos^2k}
{((nu)^2+(\sin k-\Lambda)^2)^2}f_1^2(k)\rho(k) \nonumber \\
&-&\int^{\pi}_{-\pi}\frac{dk}{\pi}
\frac{nu\sin k}{(nu)^2+(\sin k-\Lambda)^2}\frac{f_1^2(k)\rho(k)}{2}
+\sum_{m=1}^{\infty}\int^{\infty}_{-\infty} 
\frac{d\Lambda'}{2\pi}\Theta'
(\frac{\Lambda-\Lambda'}{u})
\frac{g_{2m}(\Lambda')\sigma_m(\Lambda')}{u} \nonumber \\
&+&\sum_{m=1}^{\infty}\int^{\infty}_{-\infty}
\frac{d\Lambda'}{2\pi}\Theta''
(\frac{\Lambda-\Lambda'}{u})\frac{g_{1m}^2(\Lambda')
\sigma_m(\Lambda')}{2u^2} \nonumber \\
&+&\lim_{\Lambda_0\rightarrow\infty}\frac{1}{48\pi u}
\sum_{m=1}^{\infty}\biggl[\frac{\Theta'((\Lambda-\Lambda_0)/u)}
{(1+\eta_m(\Lambda_0))\sigma_m(\Lambda)}
-\frac{\Theta'((\Lambda+\Lambda_0)/u)}
{(1+\eta_m(-\Lambda_0))\sigma_m(-\Lambda)}\biggr],
\label{g2}
\end{eqnarray}
\begin{eqnarray}
(1&+&{\eta'}_n(\Lambda)){\sigma'}_n(\Lambda)h_{2n}(\Lambda)=
\frac{1}{2}\frac{d}{d\Lambda}[(1+{\eta'}_n(\Lambda)){\sigma'}_n(\Lambda)
h_{1n}^2(\Lambda)] \nonumber \\
&-&\int^{\pi}_{-\pi}\frac{dk}{\pi}\frac{nu\cos k}
{(nu)^2+(\sin k-\Lambda)^2}f_2(k)\rho(k)
-\int^{\pi}_{-\pi}\frac{dk}{\pi}\frac{nu(\sin k-\Lambda)\cos^2k}
{((nu)^2+(\sin k-\Lambda)^2)^2}f_1^2(k)\rho(k) \nonumber \\
&+&\int^{\pi}_{-\pi}\frac{dk}{\pi}
\frac{nu\sin k}{(nu)^2+(\sin k-\Lambda)^2}\frac{f_1^2(k)\rho(k)}{2}
+\sum_{m=1}^{\infty}\int^{\infty}_{-\infty}
\frac{d\Lambda'}{2\pi}\Theta'
(\frac{\Lambda-\Lambda'}{u})
\frac{h_{2m}(\Lambda'){\sigma'}_m(\Lambda')}{u} \nonumber \\
&+&\sum_{m=1}^{\infty}\int^{\infty}_{-\infty}
\frac{d\Lambda'}{2\pi}\Theta''
(\frac{\Lambda-\Lambda'}{u})\frac{h_{1m}^2(\Lambda')
{\sigma'}_m(\Lambda')}{2u^2} \nonumber \\
&+&\lim_{\Lambda_0\rightarrow\infty}\frac{1}{48\pi u}
\biggl[\frac{\Theta'((\Lambda-\Lambda_0)/u)}
{(1+{\eta'}_m(\Lambda_0)){\sigma'}_m(\Lambda)}
-\frac{\Theta'((\Lambda+\Lambda_0)/u)}
{(1+{\eta'}_m(-\Lambda_0)){\sigma'}_m(-\Lambda))}\biggr],
\label{h2}
\end{eqnarray}
where $\Theta'(x)$ and $\Theta''(x)$ are, respectively, 
the first and second derivative of $\Theta(x)$.
This gives our starting equations for the following discussions.
Using these equations, we shall investigate finite-size effects 
to obtain the Drude weight at finite temperatures.

\section{Drude weight at finite temperatures}

Here we derive the expression for the Drude weight 
at finite temperatures using the formulation in the previous section.
The Drude weight at finite temperatures is given by the second 
derivative of the energy spectrum  with respect 
to the AB flux $\Phi$\cite{kohn},
\begin{equation}
D=\frac{L}{2}\biggl\langle\frac{d^2 E_n}{ d \Phi^2}
\biggr\rangle\biggl\vert_{\phi=0}\biggl..\label{dru}
\end{equation}  
Here $\langle \cdot\cdot\cdot\rangle$ is the thermal average
for a canonical ensemble. 
Note that the above Drude weight $D$ is different from 
the second derivative of 
the free energy, which denotes a Meissner fraction\cite{gs}.

We now wish to evaluate $D$ from the finite-size corrections
to the energy. To this end, we first write down the total energy,
\begin{eqnarray}
\frac{E}{L}&=&-\sum_{j=1}^{N-2M'}(\cos k_j+\mu_0 H+A)
+\sum_{n=1}^{\infty}\sum_{\alpha}4{\rm Re}
\sqrt{1-({\Lambda'}_{\alpha}^n-inu)^2} \nonumber \\
&&+2\mu_0H\sum_{n=1}^{\infty}nM_n-2A\sum_{n=1}^{\infty}nM_n'.
\label{ene1}
\end{eqnarray}
We then expand the energy in powers of $1/L$
using eq.(\ref{rap}),
\begin{equation}
\frac{E}{L}=E_0+\frac{E_1}{L}
+\frac{E_2}{L^2}.
\end{equation}
The first-order correction term $E_1$ is given by
\begin{equation}
E_1=2\int^{\pi}_{-\pi}dk\sin k f_1(k)\rho(k)
+\sum_{n=1}^{\infty}\int^{\infty}_{-\infty}d\Lambda4{\rm Re}
\frac{-(\Lambda-inu)}{\sqrt{1-(\Lambda-inu)^2}}h_{1n}(\Lambda)
{\sigma'}_n(\Lambda).
\label{ene2}
\end{equation}
Differentiating eqs.(\ref{tba4})-(\ref{tba6})
with respect to rapidities, and 
substituting them into eq.(\ref{ene2}), we easily find that
$E_1=0$. We now consider the second-order term which is,
\begin{eqnarray}
E_2&=&2\int^{\pi}_{-\pi}dk\sin k f_2(k)\rho(k)
+2\int^{\pi}_{-\pi}dk \cos k \frac{f_1^2(k)\rho(k)}{2} \nonumber \\
&+&\sum_{n=1}^{\infty}\int^{\infty}_{-\infty}d\Lambda
4{\rm Re}\frac{-(\Lambda-inu)}{\sqrt{1-(\Lambda-inu)^2}}
h_{2n}(\Lambda){\sigma'}_n(\Lambda) \nonumber \\
&+&\sum_{n=1}^{\infty}\int^{\infty}_{-\infty}d\Lambda
4{\rm Re}\biggl[\frac{-1}{(1-(\Lambda-inu)^2)^{3/2}}\biggr]
\frac{h_{1n}^2(\Lambda){\sigma'}_n(\Lambda)}{2}.
\end{eqnarray}
Using eqs.(\ref{tba4})-(\ref{h2}), we can rewrite 
this expression as,
\begin{eqnarray}
E_2&=&\frac{T}{2}\int^{\pi}_{-\pi}dk
\frac{\rho(k)f_1^2(k)}{\zeta(k)(1+\zeta(k))}
\biggl(\frac{d\zeta(k)}{dk}\biggr)^2
+\frac{T}{2}\sum_{n=1}^{\infty}\int^{\infty}_{-\infty}
d\Lambda\frac{\sigma_n(\Lambda)g_{1n}^2(\Lambda)}
{\eta_n(\Lambda)(1+\eta_n(\Lambda))}
\biggl(\frac{d\eta_n(\Lambda)}{d\Lambda}\biggr)^2 \nonumber \\
&+&\frac{T}{2}\sum_{n=1}^{\infty}\int^{\infty}_{-\infty}
d\Lambda\frac{{\sigma'}_n(\Lambda)h_{1n}^2(\Lambda)}
{{\eta'}_n(\Lambda)(1+{\eta'}_n(\Lambda))}
\biggl(\frac{d{\eta'}_n(\Lambda)}{d\Lambda}\biggr)^2.
\end{eqnarray}
Note that the dependence on the AB flux appears only through
$f_1(k)$, $g_{1n}(\Lambda)$, and $h_{1n}(\Lambda)$.
We can obtain the Drude weight by differentiating 
$E_2$ twice with respect to $\Phi$.
We see from eqs.(\ref{f1})-(\ref{h1}) that the equations for
$d f_1/d\Phi$, $d g_{1n}/d\Phi$ and $d h_{1n}/d\Phi$
do not depend on $\Phi$.
Thus, we have
\begin{equation}
\frac{d^2 f_1}{d\Phi^2}=\frac{d^2 g_{1n}}{d\Phi^2}
=\frac{d^2 h_{1n}}{d\Phi^2}=0.
\end{equation} 
Then the Drude weight is given by
\begin{eqnarray}
D&=&\frac{1}{2}\frac{d^2E_2}{d\Phi^2}
\biggl\vert_{\Phi=0}\biggr. \nonumber \\
&=&\frac{1}{2}\int^{\pi}_{-\pi}dk
\biggl\{(1+\zeta(k))\rho(k)\frac{d f_1}{d \Phi}\biggr\}^2
\frac{d}{dk}\biggl(\frac{-1}{1+e^{\kappa(k)/T}}\biggr)
\frac{1}{(1+\zeta(k))\rho(k)}\frac{d\kappa(k)}{dk} \nonumber \\
&+&\frac{1}{2}\sum_{n=1}^{\infty}\int^{\infty}_{-\infty}d\Lambda
\biggl\{(1+\eta(\Lambda))\sigma(\Lambda)\frac{d g_{1n}}{d \Phi}
\biggr\}^2
\frac{d}{d\Lambda}
\biggl(\frac{-1}{1+e^{\varepsilon_n(\Lambda)/T}}\biggr)
\frac{1}{(1+\eta(\Lambda))\sigma(\Lambda)}
\frac{d\varepsilon_n(\Lambda)}{d\Lambda} \nonumber \\
&+&\frac{1}{2}\sum_{n=1}^{\infty}\int^{\infty}_{-\infty}d\Lambda
\biggl\{(1+{\eta'}(\Lambda)){\sigma'}(\Lambda)
\frac{d h_{1n}}{d \Phi}\biggr\}^2
\frac{d}{d\Lambda}
\biggl(\frac{-1}{1+e^{{\varepsilon'}_n(\Lambda)/T}}\biggr)
\frac{1}{(1+{\eta'}(\Lambda)){\sigma'}(\Lambda)}
\frac{d{\varepsilon'}_n(\Lambda)}{d\Lambda}.
\label{dru2}
\end{eqnarray}
Here we have used the conventional notations, 
$\kappa(k)\equiv T\ln\zeta(k)$, 
$\varepsilon_n(\Lambda)\equiv T\ln\eta_n(\Lambda)$, and
${\varepsilon'}_n(\Lambda)\equiv T\ln{\eta'}_n(\Lambda)$. 
In order to simplify the expression, 
it is convenient to define the following quantities,
\begin{eqnarray}
\xi_c(k)&\equiv&2\pi(1+\zeta(k))\rho(k)\frac{d f_1}{d \Phi}, \\
\xi_{sn}(\Lambda)&\equiv&2\pi(1+\eta(\Lambda))\sigma(\Lambda)  
\frac{d g_{1n}}{d \Phi},  \\
\xi_{bn}(\Lambda)&\equiv&2\pi(1+{\eta'}(\Lambda)){\sigma'}(\Lambda)
\frac{d h_{1n}}{d \Phi}, \\
2\pi v_c(k)&\equiv&\frac{1}{(1+\zeta(k))\rho(k)}
\frac{d\kappa(k)}{dk}, \\
2\pi v_{sn}(\Lambda)&\equiv&\frac{1}{(1+\eta(\Lambda))\sigma(\Lambda)}
\frac{d\varepsilon_n(\Lambda)}{d\Lambda}, \\
2\pi v_{bn}(\Lambda)&\equiv&
\frac{1}{(1+{\eta'}(\Lambda)){\sigma'}(\Lambda)}
\frac{d{\varepsilon'}_n(\Lambda)}{d\Lambda}.
\end{eqnarray}
These quantities have simple physical meanings:
$\xi_c$, $\xi_{sn}$ and $\xi_{bn}$ correspond to the dressed charges
generalized to finite temperature.
$v_c$, $v_{sn}$, and $v_{bn}$ are the velocities for 
charge excitations, spin excitations, and bound states, 
respectively.  Consequently, we end up with the 
simple formula for the Drude weight expressed in terms 
of these quantities,
\begin{eqnarray}
D&=&\int^{\pi}_{-\pi}\frac{dk}{4\pi}
\frac{d}{dk}\biggl(\frac{-1}{1+e^{\kappa(k)/T}}\biggr)
\xi_c^2(k)v_c(k)
+\sum_{n=1}^{\infty}\int^{\infty}_{-\infty}\frac{d\Lambda}{4\pi}
\frac{d}{d\Lambda}
\biggl(\frac{-1}{1+e^{\varepsilon_n(\Lambda)/T}}\biggr)
\xi_{sn}^2(\Lambda)v_{sn}(\Lambda) \nonumber \\
&+&\sum_{n=1}^{\infty}\int^{\infty}_{-\infty}
\frac{d\Lambda}{4\pi}\frac{d}{d\Lambda}
\biggl(\frac{-1}{1+e^{{\varepsilon'}_n(\Lambda)/T}}\biggr)
\xi_{bn}^2(\Lambda)v_{bn}(\Lambda).
\label{dru3}
\end{eqnarray}
Note that at finite temperatures not only the charge degrees of 
freedom but also the spin degrees of freedom contribute
to the Drude weight. The above formula is the one of 
our main results  in this paper.

To conclude this section, we check that the above formula 
reproduces the known results\cite{schulz,kaya} 
  by taking zero temperature limit
$T\rightarrow 0$.
Since $\varepsilon_n(\Lambda)>0$ for $n=2,3,...$ and 
${\varepsilon'}_n(\Lambda)\geq 0$ for $n=1,2,3...$, 
the contributions from spin excitations with $n>1$ and
bound states to the Drude weight vanish for $T\rightarrow 0$.
Moreover from eqs.(\ref{f1}) and (\ref{g1}) 
we have $\xi_{s1}(\pm B)=0$ 
for $T\rightarrow 0$ where $\pm B$ is zeros of 
$\varepsilon_{1}(\Lambda)$.
Thus only the contribution from the charge degrees of freedom
to the Drude weight survives,
\begin{equation}
D=\int d\kappa \sum_{\mbox{zeros of } \kappa}
\delta(\kappa)\frac{\xi_c^2(k)v_c(k)}{4\pi}=\frac{K_cv_c}{\pi}.
\end{equation}
Here $K_c=\xi_c^2(k_0)/2$, and $v_c=v_c(k_0)$ 
with $\kappa(\pm k_0)=0$.
Then we reproduce the well-known result for zero temperature.

\section{Low temperature expansion}

\subsection{Case of half-filling}

In this section, we explicitly derive the temperature dependence of 
the Drude weight in the case of  half-filling at low temperatures.
In this case, the system is in the Mott insulating state
with the charge excitation gap. Thus we immediately 
find $D=0$ at zero temperature.
However, at finite temperatures it can have finite values as we will see
momentarily. We consider the case that $2u-A\geq 0$, 
$\mu_0 H<2(\sqrt{1+u^2}-u)$, and
$2-\mu_0 H-A\leq 0$, {\it i.e.}
the charge excitation is gapful, whereas the spin excitation 
is gapless.  In order to obtain the temperature dependence of 
the Drude weight, we need to take a thermal average over
a canonical ensemble and consider the temperature dependence 
of the chemical potential $A$. However, in the presence of 
the Mott-Hubbard gap,  the temperature dependence of $A$ appears only 
through that of the Mott-Hubbard gap,
which gives a subleading contribution to the temperature
dependence of the Drude weight, as is shown below.
Thus we can safely ignore the temperature dependence of $A$.
  
Following Takahashi's method, we perform low-temperature 
expansions\cite{taka2}.  As a result, we find that 
at low temperatures the Drude weight is mainly controlled by
the contributions from the 
charge excitation, the  spin excitation with $n=1$ and 
the bound-state excitation with $n=1$.
At low temperatures eq.(\ref{tba4}), (\ref{tba5}) for $n=1$,
and (\ref{tba6}) for $n=1$
are rewritten as,
\begin{equation}
\kappa(k)=-2\cos k-\mu_0 H A+\int^B_{-B}\frac{d\Lambda}{\pi}
\frac{u}{u^2+(\sin k-\Lambda)^2}\varepsilon_1(\Lambda)
+C_1T^{\gamma}, \label{ltek1}
\end{equation}
\begin{equation}
\varepsilon_1(\Lambda)=2\mu_0 H-4({\rm Re}
\sqrt{1-(\Lambda-iu)^2}-u)-\int^B_{-B}\frac{d\Lambda'}{\pi}
\frac{2u}{4u^2+(\Lambda-\Lambda')^2}\varepsilon_1(\Lambda')
+C_2T^{\gamma}, \label{ltee1}
\end{equation}
\begin{equation}
{\varepsilon'}_1(\Lambda)=4u-2A-C_3T^{\frac{3}{2}}
e^{\kappa(\pi)/T}\frac{u}{u^2+\Lambda^2}
+C_4T^{\frac{3}{2}}e^{-(4u-2A)/T}, \label{lteb1}
\end{equation}
where $\gamma=2$ for $H\neq 0$ and $\gamma=3/2$ for $H=0$.
$B$ is defined by the condition $\varepsilon_1(\pm B)=0$.
In the absence of magnetic fields, $H=0$, $B\rightarrow +\infty$.
$\varepsilon_1(\Lambda)$ is obtained from eq.(\ref{ltee1}) 
by using the Wiener-Hopf method.
Substituting the solution into eqs.(\ref{ltek1}) 
and (\ref{lteb1}),
we obtain $\kappa(k)$ and ${\varepsilon'}_1(\Lambda)$.
The generalized dressed charges $\xi_c$, $\xi_{s1}$ and $\xi_{b1}$
are now determined by
the derivative of eqs.(\ref{f1}), (\ref{g1}), and (\ref{h1}) 
with respect to $\Phi$,
which are given in the low temperature limit,
\begin{equation}
(1+e^{\kappa(k)/T})\xi_c(k)=
1+\int^{B}_{-B}\frac{d\Lambda}{\pi}
(\xi_{s1}(\Lambda)+\xi_{b1}(\Lambda)),
\end{equation}
\begin{equation}
(1+e^{\varepsilon_1(\Lambda)/T})
\xi_{s1}(\Lambda)=\int^{\pi}_{-\pi}\frac{dk}{\pi}
\frac{u\cos k}{u^2+(\sin k-\Lambda)^2}\xi_c(k)-\int^{B}_{-B}
\frac{d\Lambda'}{\pi}\frac{2u}{4u^2+(\Lambda-\Lambda')^2}
\xi_{s1}(\Lambda).
\end{equation}
\begin{equation}
(1+e^{{\varepsilon'}(\Lambda)/T})\xi_{b1}(\Lambda)
=2+\int^{\pi}_{-\pi}\frac{dk}{\pi}
\frac{u\cos k}{u^2+(\sin k-\Lambda)^2}\xi_c(k)-\int^{B}_{-B}
\frac{d\Lambda'}{\pi}\frac{2u}{4u^2+(\Lambda-\Lambda')^2}
\xi_{b1}(\Lambda).
\end{equation}
Solving these equations we have $\xi_c(k)=1$ and
\begin{equation}
\xi_{s1}(\Lambda)\sim D_1\sqrt{T}e^{(2-\mu_0 H -A+\Delta(B))/T},
\end{equation}
\begin{equation}
\xi_{b1}(\Lambda)\sim D_2e^{-{\varepsilon'}_1(\Lambda)/T},
\end{equation}
where 
\begin{equation}
\Delta(B)\equiv\int^{B}_{-B}\frac{d\Lambda}{\pi}
\frac{u}{u^2+\Lambda^2}\varepsilon_1(\Lambda).
\end{equation}
The distribution functions for rapidities, $\rho(k)$ 
and $\sigma_1(\Lambda)$ are given by those for zero temperature.
Also, ${\sigma'}_1(\Lambda)$ is estimated as
\begin{equation}
{\sigma'}_1(\Lambda)\sim C'\sqrt{T}
e^{(2-\mu_0 H -A+\Delta(B))/T}
e^{-{\varepsilon'}_1(\Lambda)/T}.
\label{sigb}
\end{equation}
Using eqs.(\ref{dru3}) and (\ref{ltek1})-(\ref{sigb}), we 
finally end up with 
\begin{equation}
D=\frac{\sqrt{T}}{\sqrt{\pi}\tilde{\rho}}e^{-\Delta_{\rm MH}/T}
+O(Te^{2(2-\mu_0 H-A+\Delta(B))/T})
+O(T^{3/2}e^{-(4u-2A)/T}e^{(2-\mu_0 H -A+\Delta(B))/T}),
\label{drult}
\end{equation}
where $\Delta_{\rm MH}\equiv -2+\mu_0 H+A-\Delta(B)$ is nothing 
but the Mott-Hubbard gap, and
\begin{eqnarray}
\tilde{\rho}&=&\frac{1}{2\pi}-\int^{B}_{-B}\frac{d\Lambda}{\pi}
\frac{u\sigma_0(\Lambda)}{u^2+\Lambda^2}, \\
\sigma_0(\Lambda)&=&\int^{\pi}_{-\pi}
\frac{dk}{8\pi u\cosh \frac{\pi(\Lambda-\sin k)}{2u}}.
\end{eqnarray}
Here the first term of eq.(\ref{drult}), 
which is most dominant, comes 
from the charge degrees of freedom, whereas the 
second and third terms are 
the contributions from spin degrees of freedom and bound states, 
respectively. As seen from the above equations, the Drude weight 
vanishes exponentially at half-filling at low temperatures,
reflecting the presence of the Mott-Hubbard gap. 

\subsection{Case away from half-filling}

We next consider the case away from half-filling 
in the absence of magnetic fields, {\it i.e.}
$\kappa(\pi)>0$.
We first estimate the contribution from the 
charge degrees of freedom. Since we are concerned with a canonical 
ensemble, we must take into account the temperature dependence of 
the chemical potential $A$ or $k_0$ 
in the case that the  number of electrons
\begin{equation}
\frac{N}{L}=\int^{\pi}_{-\pi}dk\rho(k)+\sum_{n=1}^{\infty}
2n\int^{\infty}_{-\infty}d\Lambda{\sigma'}_n(\Lambda),
\label{num}
\end{equation}
is fixed. At low temperatures eq.(\ref{num}) is approximated by
\begin{eqnarray}
\frac{N}{L}&\approx&\int^{\pi}_{-\pi}dk
\frac{\rho_0(k)}{1+e^{\kappa(k)/T}}  \nonumber \\
&\approx&\frac{k_0}{\pi}+2\int^{\infty}_{-\infty}
\frac{\Lambda}{\pi}\tan^{-1}\frac{\sin k_0-\Lambda}{U}
\sigma_1(\Lambda)
+\frac{\pi^2T^2}{3}\frac{\partial }{\partial \kappa}
\biggl(\frac{\rho_0(k)}{\kappa'(k)}\biggr)\biggl\vert_{k=k_0}
\biggr.,
\end{eqnarray}
where $\rho_0(k)$ is the distribution function for $k$ at $T=0$.  
Thus we have the temperature dependence of $k_0$,
\begin{equation}
\delta k_0 \equiv k_0-\tilde{k}_0=
-\frac{\pi^2T^2}{6\rho_0(\tilde{k}_0)}
\frac{\partial }{\partial \kappa}
\biggl(\frac{\rho_0(k)}{\kappa'(k)}\biggr)
\biggl\vert_{k=\tilde{k}_0}\biggr..
\label{k0}
\end{equation}
Here $\tilde{k}_0$ is $k_0$ at $T=0$.
Then using eq.(\ref{dru3}), we obtain 
the contribution from the charge degrees of freedom to 
the Drude weight at low temperatures,
\begin{eqnarray}
D_{\rm charge}&\approx&\frac{\xi^2_c(k_0)v_c(k_0)}{2\pi}
\frac{\pi T^2}{12}\frac{\partial^2}{\partial \kappa^2}
(\xi_c^2(k)v_c(k))\biggl\vert_{k=k_0}\biggr. \nonumber \\
&\approx&\frac{K_cv_c}{\pi}+\frac{\delta k_0}{2\pi}
\frac{\partial }{\partial k_0}
(\xi_c^2(k_0)v_c(k_0))\biggl\vert_{k_0=\tilde{k}_0}\biggr.
+\frac{\pi T^2}{12}
\frac{\partial^2}{\partial \kappa^2}
(\xi_c^2(k)v_c(k))\biggl\vert_{k=\tilde{k}_0}\biggr. \nonumber \\
&=&\frac{K_cv_c}{\pi}+ CT^2,
\label{drucha}
\end{eqnarray}
where the coefficient of the quadratic term is 
\begin{eqnarray}
C=\frac{\pi}{12}\biggl[\frac{\partial^2}{\partial \kappa^2}
(\xi_c^2(k)v_c(k))\biggl\vert_{k=\tilde{k}_0}\biggr. 
-\frac{1}{\rho_0(\tilde{k}_0)}\frac{\partial }{\partial \kappa}
\biggl(\frac{\rho_0(k)}{\kappa'(k)}\biggr)
\biggl\vert_{k=\tilde{k}_0}\biggr.
\frac{\partial }{\partial k_0}
(\xi_c^2(k_0)v_c(k_0))\biggl\vert_{k_0=\tilde{k}_0}\biggr.
\biggr].
\end{eqnarray}
Here $\xi_c(k)$ and $v_c(k)$ are calculated at $T=0$.

In a similar manner, we can evaluate the contributions 
from the spin degrees of freedom as well as from the bound states.
As a consequence we find that
they are subdominant compared to the contribution
from the charge degrees of freedom at low temperatures.
Therefore the leading term of the Drude weight is given by
eq.(\ref{drucha}).
We expect that the expression (\ref{drucha}) 
and its temperature dependence $\sim D(0)+CT^2$ 
may be general for all integral models
with massless excitations.
	
\section{Summary}

By using the Bethe ansatz solution, we have obtained the formula 
for the Drude weight of the Hubbard model at
finite temperatures. The present general formulation is 
not restricted to the Hubbard model, but also applicable to
any other integrable models.
We have then performed low-temperature expansions 
both in the case of half-filling as well as away from half-filling.
In the case of half-filling, the Drude weight 
decreases exponentially, as the temperature is lowered,
reflecting the presence of the Mott-Hubbard gap.
In the case away from half-filling,
it behaves like $\sim D(0)+CT^2$, with the coefficient $C$
expressed in terms of the velocity for charge excitations,
the dressed charge and their derivatives.  Although 
the essential properties of the Drude weight can be 
seen through the present low-temperature expansion, 
it is interesting to obtain its full-temperature 
dependence  by solving the integral equations numerically, 
which should be done in the future study.


\acknowledgments{}
This work was partly supported by a Grant-in-Aid from the Ministry
of Education, Science, Sports and Culture, Japan.


                                                                    
\end{document}